\documentclass[aps,prl,twocolumn,showpacs,preprintnumbers,amsmath,amssymb]{revtex4-1}

\usepackage{graphicx}
\usepackage{dcolumn}
\usepackage{bm}
\begin{document}


\title{Correlation between nuclear temperatures and symmetry energy in sub-saturation}

\author{Fan Zhang$^{1}$}\email[]{Corresponding author: zhangfan@czc.edu.cn}
\author{Kai-Lei Wang $^{1}$}
\author{Li-Ding Jin $^{1}$}
\author{Zhu-Li Zhang $^{1}$}
\author{Hui-Xiao Duan $^{1}$}
\author{Cheng Li $^{2,3}$}\email[]{Corresponding author: licheng@mail.bnu.edu.cn}



\affiliation{
$^{1}$Department of Physics, Changzhi University, Changzhi 046011, PR China \\
$^{2}$Department of Physics, Guangxi Normal University, Guilin 541004, PR China \\
$^{3}$Guangxi Key Laboratory of Nuclear Physics and Technology, Guilin 541004, PR China \\
}


\date{\today}

\begin{abstract}
A study of the correlation between nuclear temperatures and symmetry energy is presented for heavy-ion collisions at intermediate energies via the isospin-dependent quantum molecular-dynamics model. It is found that different symmetry energy parameters will change the density and kinetic energy distribution of the hot nuclei. More importantly, nuclear temperatures that are based on kinetic energy properties can be used to study symmetry energy information.
\end{abstract}


\maketitle
\section{ INTRODUCTION}
Nuclear symmetry energy, which has been a research focus of nuclear physics for many years, governs important properties of nuclei and neutron stars \cite{fus1}. Compared to regions with saturation density, the large uncertainties in symmetry energy exist in low- and high-density regions. To reduce the uncertainties regarding symmetry energy in non-saturated density regions, many investigations have been undertaken; at sub-saturation densities these include isotopic scaling \cite{fus2,fus3}, isospin fractionation \cite{fus4,fus5}, pre-equilibrium single and double neutron-proton ratios \cite{fus6,fus7,fus8,fus9} and isobaric ratio of various species \cite{fus10} etc. However, the high-density behavior of symmetry energy has been studied using methods including collective and elliptic flows \cite{fus11,fus12,fus13}, neutron-proton ratios of free nucleons \cite{fus14,fus15}, $\pi^{-}/\pi^{+}$ \cite{fus16,fus17}, $K^{+}/K^{0}$ \cite{fus18} etc. Numerous efforts have been directed towards constraining symmetry energy, but further studies are needed to improve the accuracy of the constraint on the symmetry energy at sub- and supra-saturation densities.

Recent experimental and theoretical studies show that nuclear temperatures have isospin dependence \cite{fus19,fus20,fus21,fus22,fus23,fus24}.
There are two main factors contributing to this phenomenon: the Coulomb interaction and symmetry energy. Based on Landau theory, McIntosh $\emph{\textrm{et al}}$ revealed a linear dependence of temperature on Coulomb energy and symmetry energy \cite{fus20}. Therefore, if one could select appropriate hot nuclei and reduce the effects of the Coulomb interaction, one might use the nuclear temperature isospin dependence to study symmetry energy. In this work, we focus on symmetry energy at sub-saturation densities and examine the correlation between nuclear temperatures and symmetry energy in low-density regions.

\section{ MODEL AND METHODS}
We attempt to study the correlation between nuclear temperatures and symmetry energy in a low-density region via the isospin-dependent quantum molecular dynamics ($\textrm{IQMD}$) model \cite{fus25,fus26,fus27,fus28} incorporating the statistical decay model GEMINI \cite{fus29}. To better connect the two models, we needed a dynamical model to describe the intermediate-mass-fragment (IMF) emission. When the maximum fragment excitation energy is less than a certain value $E_{\textrm{stop}}$, the dynamical simulation will stop and the statistical decay model will complete the decay of pre-fragments. The value of $E_{\textrm{stop}}$
corresponds to the threshold energy for IMF emission. In this work, $E_{\textrm{stop}}$ = 2 $\textrm{MeV}/\textrm{nucleon}$. Using this value, the experimental data for IMF production can be described very well \cite{fus27}.

In the present model, the Hamiltonian $H$ is expressed as
\begin{equation}
H=\tau+U_{Coul}+\int V(\rho)d\emph{r},
\end{equation}
where $\tau$ is the kinetic energy and $U_{Coul}$ is the Coulomb potential energy. $V(\rho)$ is the nuclear potential energy density function, which is written as
\begin{eqnarray}
V(\rho)=&&\frac{\alpha}{2}\frac{\rho^{2}}{\rho_{0}}+
\frac{\beta}{\gamma+1}\frac{\rho^{\gamma+1}}{\rho_{0}^{\gamma}}
+\frac{\textrm{\textrm{g}}_{sur}^{iso}}{2}\frac{(\nabla\rho_{n}-\nabla\rho_{p})^{2}}{\rho_{0}}\\
&&\nonumber
+\frac{\textrm{g}_{sur}}{2}\frac{(\nabla\rho)^{2}}{\rho_{0}}+
\textrm{g}_{\tau}\frac{\rho^{8/3}}{\rho_{0}^{5/3}}
+\frac{C_{sym}}{2}\left(\frac{\rho}{\rho_{0}}\right)^{\gamma_{i}}\rho\delta^{2}.
\end{eqnarray}
The parameters used in this study are $\alpha$ = -168.40 MeV, $\beta$ = 115.90 MeV, $\gamma$ = 1.50, $g_{sur}$ = 92.13 MeV $\textrm{fm}^{2}$, $g_{sur}^{iso}$ = -6.97 MeV $\textrm{fm}^{2}$, $C_{sym}$ = 38.13 MeV, and $g_{\tau}$ = 0.40 MeV. The corresponding compressibility is 271 MeV \cite{fus30}. In this work, we used three symmetry parameters $\gamma_{i}$ =0.5, 1.0 and 2.0 which correspond to the soft, linear and hard symmetry energy, respectively.

To study the energy distribution of the hot nuclei, we must find the hot nuclei in the early stage of the reactions from the phase space. The hot nuclei can be selected by the relative distance ($\textrm{R}_{\textrm{p}}$) among the nucleons. If the relative distance between the two nucleons is smaller than $\textrm{R}_{\textrm{p}}$, they can be recognized to belong one cluster. In this work, $\textrm{R}_{\textrm{p}}$ = 3 fm, which is the typical value of nuclear force scope. To further select the equilibrated projectile spectator, spherical spectators are selected by the ratio of parallel to transverse quantities:

\begin{equation}
Q_{\textrm{shape}}=\frac{2\sum_{i}p_{zi}^{2}}
{\sum_{i}(p_{xi}^{2}+p_{yi}^{2})},
\end{equation}
where $p_{xi}$, $p_{yi}$, and $p_{zi}$ are the momentum components of the $\emph{i}$th nucleon along the x, y, and the z axes in the center-of-mass frame of the projectile spectator. If the $Q_{\textrm{shape}}$ value of the spectator satisfies -0.3 $\leqslant$ $log_{10}(Q_{\textrm{shape}})$ $\leqslant$ 0.3, the spectator is a candidate that may be used to study thermodynamic properties. To use the spectator to study nuclear temperatures, the mass and neutron-proton ratio requirement must also be met, because a large mass and neutron-proton ratio range will affect the nuclear temperatures isospin effects measurement \cite{fus31}.
To reduce the effects of mass and neutron-proton ratio on nuclear temperature measurement, the mass and neutron-protons ratio range of hot nuclei should be 185 $\leq$ A $\leq$ 195 and 1.3 $\leq$ N/Z $\leq$ 1.4, respectively.

When the hot nuclei are confirmed, the excitation energy and temperature of the hot nuclei can be calculated.
The excitation energy $E^{*}$ of the hot nuclei is calculated by
\begin{equation}
E^{*}=\tau+V-B
\end{equation}
where $\tau$ and $V$ are kinetic energy and potential energy of the hot nuclei, respectively. $B$ is the binding energy of the hot nuclei at the ground state.
The temperatures of the spectator are calculated by the momentum quadrupole temperature \cite{fus32}:
\begin{equation}
\langle \sigma_{xy}^{2}\rangle=4m^{2}T^{2},
\end{equation}
where m is the probe particle mass and $\langle\sigma_{xy}^{2}\rangle$ the variance of the momentum quadrupole.

\section{ RESULTS AND DISCUSSION}
\begin{figure}
  \centering
  \includegraphics[width=8.6cm]{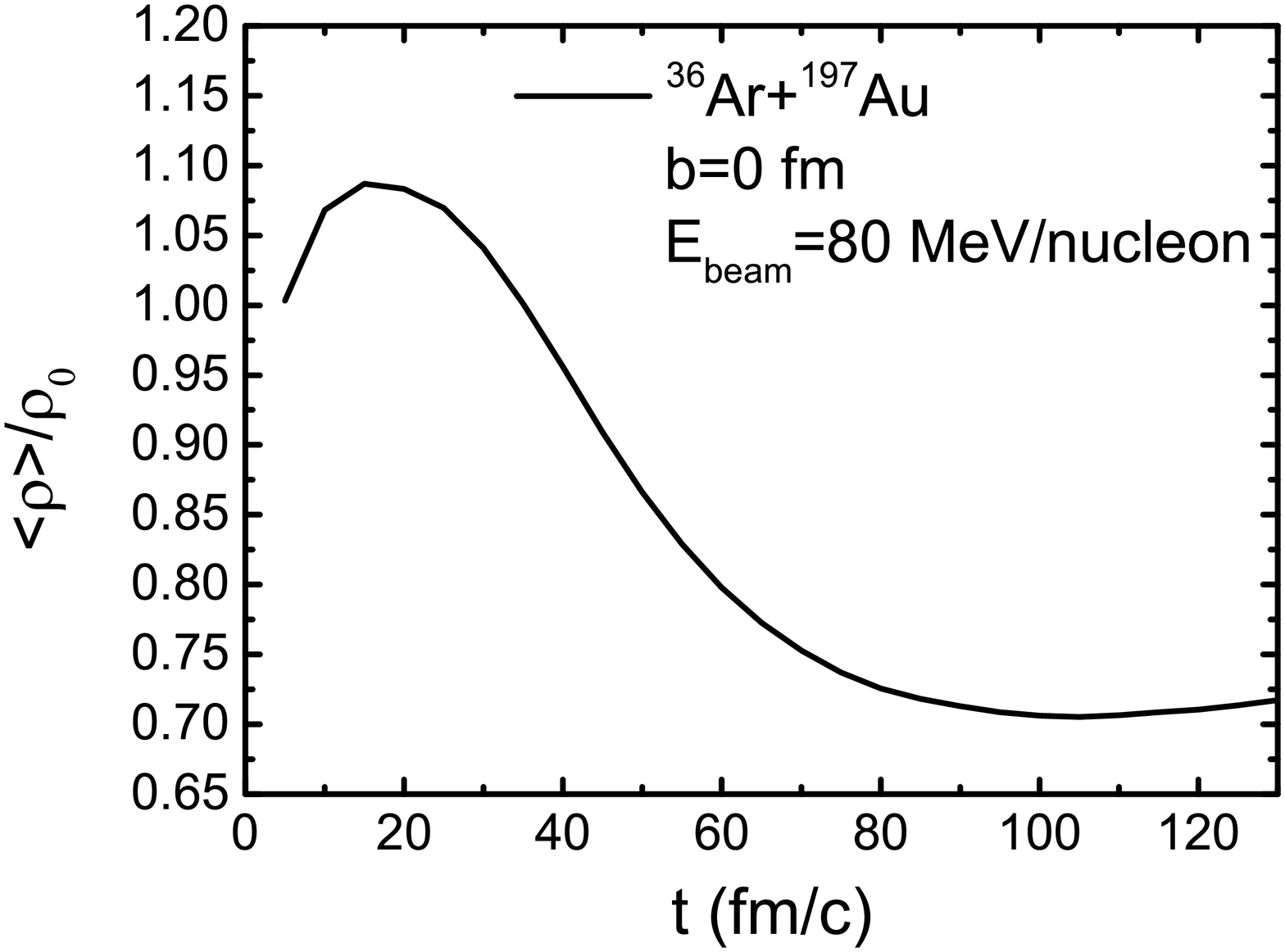}\\
  \caption{The time evolution of the largest cluster average density. }
\end{figure}
Figure 1 shows the time evolution of the largest cluster average density. In it, the reaction system is $^{36}\textrm{Ar}$ + $^{197}\textrm{Au}$ at 80 MeV/nucleon with central collisions. It is worth mentioning that the largest cluster is not the same as the hot nuclei that are used to calculate energy distribution and nuclear temperature. When the $Q_{\textrm{shape}}$, mass and N/Z values meet the requirements of the hot nuclei, the largest cluster is a hot nuclei, and will be used in the next step of this study.
It can be seen from Fig. 1 that the largest cluster average density reaches the maximum value of approximate 20 fm/c. At this moment, the reaction system reaches maximum compression. After 20 fm/c, the largest cluster average density decreases with reaction time, which is caused by the expansion of the reaction system. When the momentum distribution of the largest cluster reaches isotropy (approximate 110 fm/c \cite{fus22}), the largest cluster is a candidate for a hot nucleus which can be used to study nuclear temperature.
The average density of the hot nuclei candidates is in the sub-saturation density  region and is approximate 0.7$\rho_{0}$. If one studies the temperatures of the hot nuclei, the nuclear temperatures should carry the information about nuclear symmetry energy at sub-saturation densities.
\begin{figure}
  \centering
  \includegraphics[width=8.6cm]{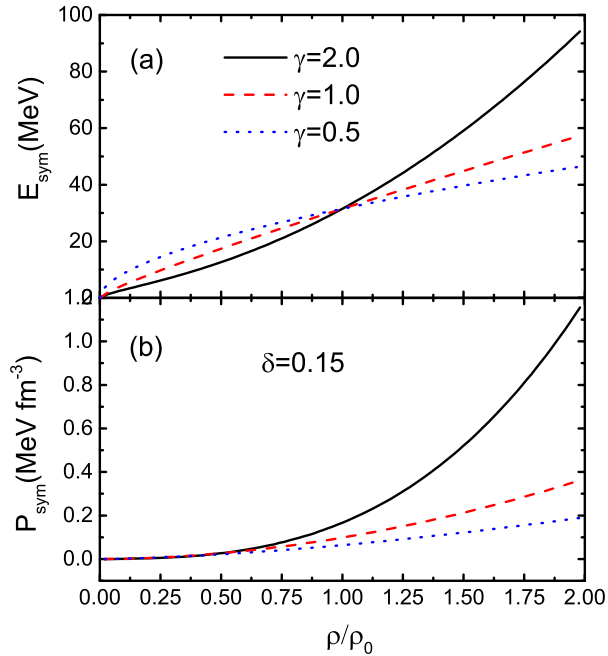}\\
  \caption{Symmetry energy (a) and pressure of symmetry energy (b) as a function of density for an isospin asymmetry of $\delta$ = 0.15 and $\gamma_{i}$ parameter of 0.5, 1.0 and 2.0, respectively. }
\end{figure}
It can be seen from Fig. 1 that the average density of the largest cluster is below saturation density (0.7$\rho_{0}$ $<$ $\rho$ $<$ 0.9$\rho_{0}$) in the formation process (50 fm/c $<$ t ). In this density region, the pressure caused by symmetry energy is positive. The role of symmetry energy is to make the system easier to expand.

The pressure of the symmetry energy can be written as
\begin{equation}
P_{sym}=\rho^{2} \left(\frac{\partial \emph{e}_{sym}}{\partial \rho}\right)_{T,\delta=constant},
\end{equation}
where $\emph{e}_{sym}$ is $E_{sym}\delta^{2}$. In the present work, $\delta$ = 0.15 which corresponds to hot nuclei neutron-proton asymmetry. Since the pressure increases with the slope of symmetry energy. It is seen from Fig. 2(b) that the hard ($\gamma_{i}$ = 2.0) symmetry energy leads to a larger pressure than the soft symmetry energy
($\gamma_{i}$ = 0.5) at densities above 0.7$\rho_{0}$.
The hot nuclei with hard symmetry energy will be much easier to expand.
Therefore, the density of the hot nuclei which use hard symmetry energy should be the lowest.
To show the effects of symmetry energy on the hot nuclei properties, the density and energy distribution are compared in Fig. 3 and Fig.4 at 110 fm/c.

The density versus excitation energy of the hot nuclei is shown in Fig. 3.
To obtain hot nuclei with different excitation energies, the reaction systems $^{36}\textrm{Ar}$ $+$ $^{197}\textrm{Au}$ at 70, 75 and 80 MeV/nucleon with different symmetry energy parameters are used.
It can be seen from Fig. 3 that the average density is the lowest for the hard symmetry energy, which supports the above reasoning.
\begin{figure}
  \centering
  \includegraphics[width=8.6cm]{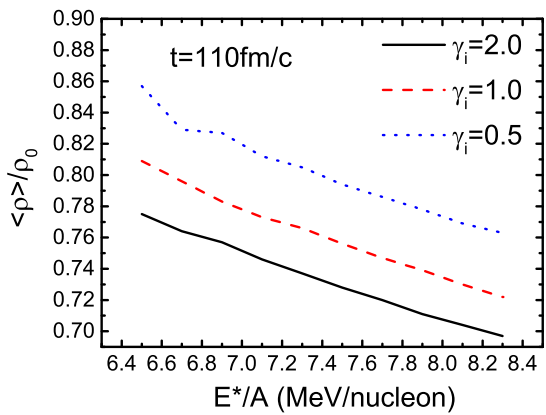}\\
  \caption{The hot nuclei average density as a function of excitation energy for different asymmetry parameter. }
\end{figure}
We can also see from Fig. 3 that the density decreases with increasing excitation energy. Using the same symmetry energy the hot nuclei will expand more easily with higher excitation energy. Similar result has been found by Wuenschel $et$ $al$ \cite{fus32}.

To further investigate the correlation between symmetry energy and nuclear temperature, the energy distribution of the hot nuclei is shown in Fig.4, in which the average total energy $E_{\textrm{tot}}$ of the hot nuclei is divided into three parts. They are average potential energy $E_{\textrm{pot}}$, collective kinetic energy $E_{\textrm{coll}}$ and intrinsic kinetic energy $E_{\textrm{int}}$,
\begin{equation}
E_{\textrm{tot}}=E_{\textrm{pot}}+E_{\textrm{coll}}+E_{\textrm{int}}.
\end{equation}
$E_{\textrm{int}}$ includes Fermi kinetic energy and thermal kinetic energy. The difference of Fermi motion and thermal motion come from the change of hot nuclei density, which carries symmetry energy information.

It can be seen from Fig.4 that the potential energy per nucleon is the highest for hard symmetry energy. This is due to the hot nuclei with hard symmetry energy having the lowest density (see Fig. 3).
The difference in the potential energy among the different symmetry energies is approximately 2 MeV/nucleon. Compared to the potential energy, the difference of the collective kinetic energy is weak.
The difference in $\textrm{E}_{\textrm{coll}}$ is approximately 0.2-0.6 MeV/nucleon for different excitation energies.
For the hot nuclei that have the same excitation energy, the intrinsic kinetic energy will be higher for the hot nuclei with soft symmetry energy. It can be seen from Fig. 4(a) that the difference in $E_{\textrm{int}}$ among different symmetry energies is approximately 2 MeV/nucleon; the $E_{\textrm{int}}$ value is higher for soft symmetry energy. It is worth mentioning that $E_{\textrm{int}}$ at 110 fm/c is not particle kinetic energy which is measured by experiment. Because the emitted particles need to overcome potential energy attraction.

\begin{figure}
  \centering
  \includegraphics[width=8.6cm]{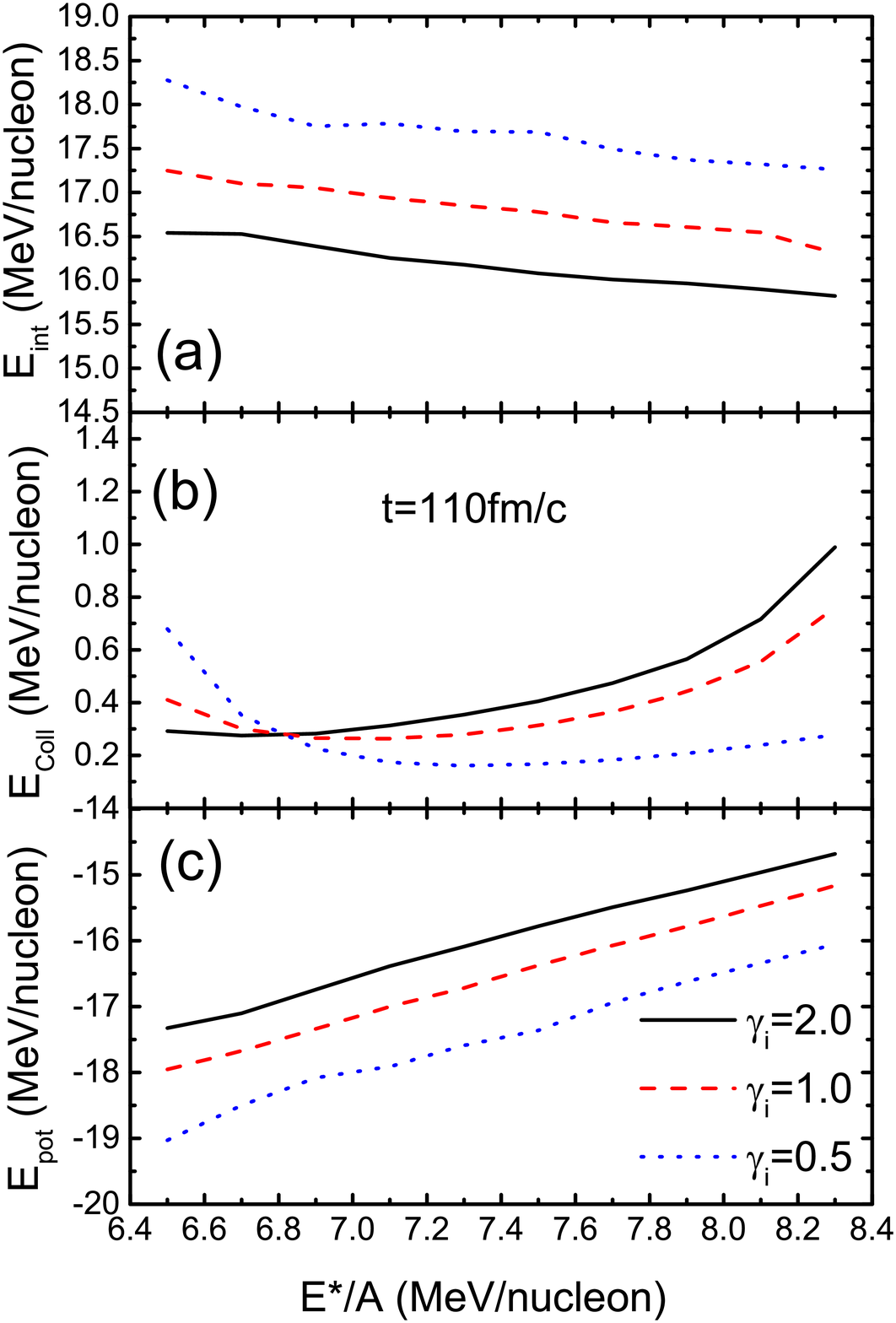}\\
  \caption{The intrinsic kinetic energy (a), collective kinetic energy (b) and potential energy (c) as a function of excitation energy. }
\end{figure}

\begin{figure}
  \centering
  \includegraphics[width=8.6cm]{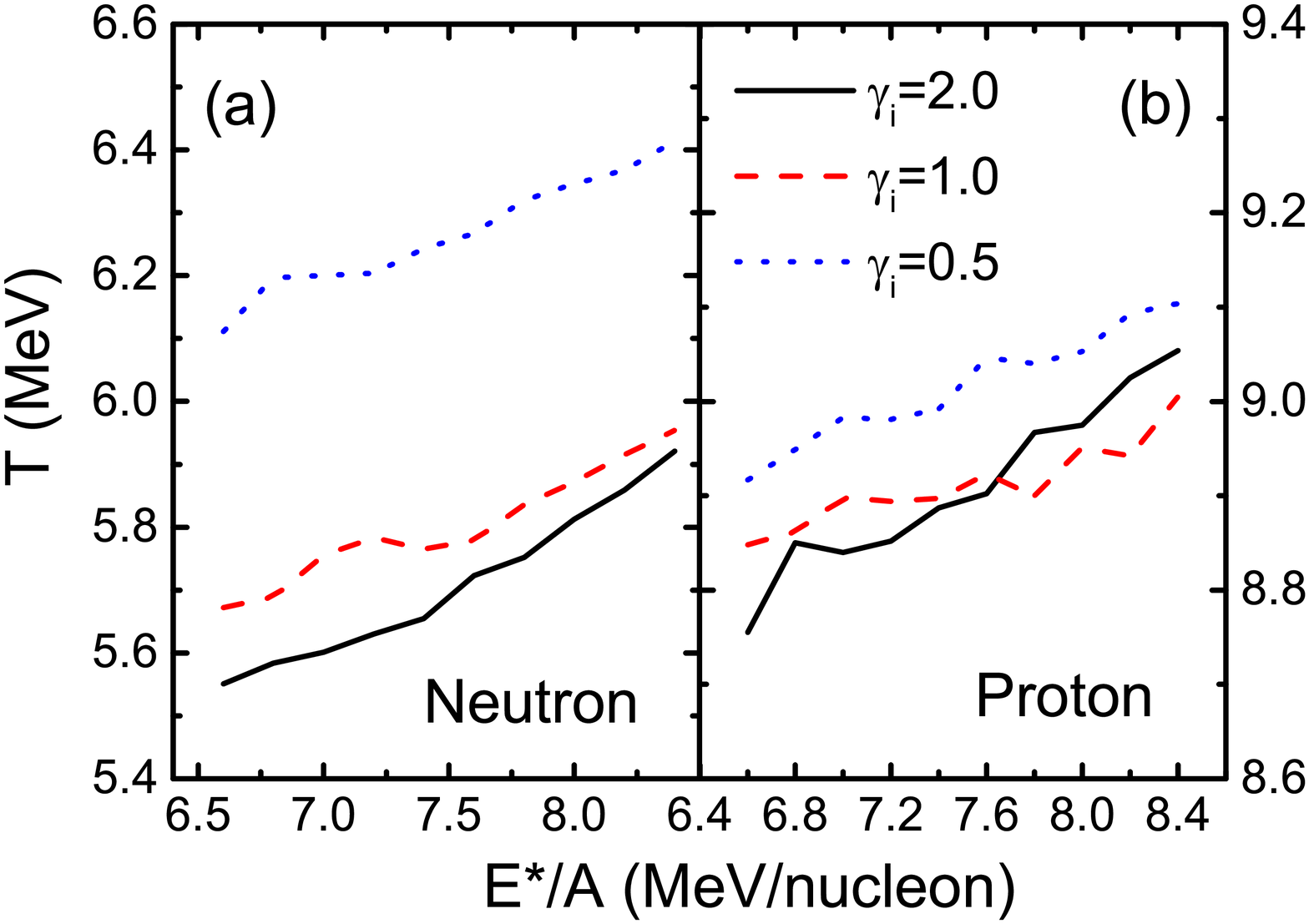}\\
  \caption{The nuclear temperatures of the hot nuclei as a function of excitation energy for different asymmetry parameter. }
\end{figure}
However, the emitted particles still carry the information of $E_{\textrm{int}}$ and reflect the difference of symmetry energy. Therefore,
the difference in $E_{\textrm{int}}$ among the different symmetry energies is expressed by particle momentum. Based on the classical Maxwell distribution, the nuclear temperature can be calculated by Eq. (5). In Fig. 5 neutrons and protons were selected as the probe particles, the yields of which are enough to satisfy statistical requirements.
It can be found from Fig. 5(a) that the softer symmetry parameter the temperatures of the hot nuclei are higher. Compared to neutrons, the difference of nuclear temperature among different symmetry energy is weak when protons are used [Fig. 5(b)]. This is mainly caused by Coulomb effect. After the neutrons and protons are created, the momentum of protons is changed by Coulomb force. Thermonuclear information carried by protons is affected.
Therefore, the influence of Coulomb effect should be minimized when the classical momentum quadrupole thermometer are used to extract symmetry energy information at the sub-saturation density region.

\section{ CONCLUSIONS}
In summary, we have presented the details of a study of the relation between the momentum quadrupole temperature and symmetry energy using the IQMD model. We found that by using different symmetry energies, the energy distribution and average density of the hot nuclei will be changed. More interesting is that a strong correlation exists between the 'classical' momentum quadrupole temperature and symmetry energy in the sub-saturation density region.

\vskip 10mm

\begin{center}
\textbf{ACKNOWLEDGEMENTS}
\end{center}

This work was supported by the National Natural Science Foundation of China under Grants Nos. 11847036 and 11905018.

\end{document}